\documentclass[a4paper,11pt]{article}
\usepackage{jcappub} 
\usepackage{indentfirst}
\usepackage{epsf,amsmath,graphicx,amssymb,url,hyperref,dcolumn}
\usepackage{mathtools}
\usepackage{color}
\usepackage{longtable,tabu}
\usepackage[normalem]{ulem}
\usepackage{adjustbox}
\usepackage{array}
\usepackage{xcolor}

\newcommand{\NIKHEF}{Nikhef - National Institute for Subatomic Physics, Science Park, 1098 XG Amsterdam, The Netherlands}

\newcommand{\UVA}{Institute for High-Energy Physics, University of Amsterdam, Science Park, 1098 XG Amsterdam, The Netherlands}
 
\newcommand{\GRASP}{Institute for Gravitational and Subatomic Physics, Utrecht University, Princetonplein 1, 3584 CC Utrecht, The Netherlands}

\newcommand{\UmasD}{Department of Physics, University of Massachusetts, Dartmouth, MA 02747, The USA}

\newcommand{\UoB}{School of Physics and Astronomy and Institute for Gravitational Wave Astronomy,\\University of Birmingham, Edgbaston, Birmingham, B15 9TT, United Kingdom}

\title{\boldmath Dipole Anisotropy in Gravitational Wave Source Distribution}
\author[a]{Gopal Kashyap}
\emailAdd{gplkumar87@gmail.com}
\affiliation[a]{Department of Physics, School of Advanced Sciences,
Vellore Institute of Technology, Vellore, Tamil Nadu 632014, India}
\author[b]{Naveen K. Singh}
\emailAdd{naveen.nkumars@gmail.com}
\affiliation[b]{Sir P.T. Sarvajanik College of Science, Surat -395001, Gujarat, India}
\author[c,d,e,f]{Khun Sang Phukon}
\emailAdd{ksphukon@star.sr.bham.ac.uk}
\affiliation[c]{\NIKHEF}
\affiliation[d]{\UVA}
\affiliation[e]{\GRASP}
\affiliation[f]{\UoB}
\author[c,e,g]{Sarah Caudill}
\affiliation[g]{\UmasD}
\emailAdd{scaudill@umassd.edu}
\author[h]{Pankaj Jain}
\emailAdd{pkjain@iitk.ac.in}
\affiliation[h]{Department of Space Science \& Astronomy, Indian Institute of Technology, Kanpur  208016, India}

\abstract{
 Our local motion with respect to the cosmic frame of rest is believed to be dominantly responsible for the observed dipole anisotropy in the Cosmic Microwave Background Radiation (CMBR). We study the effect of this motion on the sky distribution of gravitational 
 wave (GW) sources. We determine the resulting dipole anisotropy in GW source number counts, mass weighted number counts, which we refer to as mass intensity, 
 and mean mass per source. The mass $M$ dependence of the number density n(M) distribution of BBH is taken directly from data. We also test the anisotropy in the observable mean mass per source along the direction of the CMB dipole. The current data sample is relatively small and consistent with isotropy. The number of sources required for this test is likely to become available in near future.}

\begin{document}
\maketitle
\flushbottom

\section{Introduction} \label{sec_1}
The standard $\Lambda$CDM model of modern cosmology is based on the assumption of isotropy and homogeneity of the Universe at large distance scales, known as the Cosmological Principle. A
testable consequence of this would be an isotropic distribution of galaxies at large distance scales. There are many indications that the Universe may be isotropic at scales $\gtrsim$ 100 Mpc 
\cite{smoot1992structure, fixsen1996, wu1999large, blake2002velocity, marinoni2012scale, meegan1992spatial, scharf1999evidence, planck2020_I, planck2014}. The most used source
for the measurement of the isotropy of the Universe is the Cosmic Microwave Background Radiation (CMBR). The analysis of CMBR temperature and polarisation fluctuations using 
Planck observations shows small-scale statistical deviations which are consistent with the cosmological principle. The largest temperature anisotropy in the CMBR is the dipole which is interpreted in terms 
of our local motion with respect to the CMB rest frame and has been measured in Refs. \cite{Kogut1993,Hinshaw2009,fixsen1996,planck2014,planck2020_I,planck2020_II}. Considering
the dipole as being solely caused by our motion, the velocity is found to be (369.82$\pm$ 0.11 Km s$^{-1}$) in the direction, $l=264.021^o\pm 0.011^o$, $b=48.253\pm 0.0005^o$ in
galactic coordinates \cite{Kogut1993,Hinshaw2009,planck2020_I}. In J2000 equatorial coordinates, the direction parameters are $RA=167.9^o$, $DEC=-6.93^o$. 

Our local motion is also expected to generate a dipole anisotropy in the large scale structure \cite{ellis,Baleisis:1998,Condon:1998}. Such an anisotropy has been observed 
 in the distribution of radio sources \citep{Blake:2002,Crawford:2009,Singal:2011,Gibelyou:2012,Rubart:2013,tiwari2015} and in the diffuse
X-ray background \cite{Boughn_2002}. These studies find reasonable agreement
with the direction of CMBR dipole. However, surprisingly the dipole amplitude in radio observations is found to be much
larger than predicted \cite{Singal:2011,Gibelyou:2012,Rubart:2013,tiwari2015}. 
 A strong signal of deviation from the CMB prediction is also seen in the infrared observations \cite{Secrest:2020}. A recent study attributes these deviations, along with the observed Hubble tension, to a large scale inhomogeneity in the Universe, arising due to cosmic super-horizon perturbation modes \citep{Tiwari:2021}. 
  While these inferences may be speculative, the observation of gravitational waves provides an independent probe to study the isotropy of the Universe and test these claims.

The detection of gravitational waves by Advanced LIGO \cite{LIGO2015} and Advanced Virgo \cite{Virgo2014}  detectors revealed the existence of a detectable population of coalescing 
stellar-mass binary black holes (BBHs). Studies of gravitational wave signals from resolved, well-localized BBH mergers present another tool to probe the anisotropy of the Universe. Advanced LIGO
and Advanced Virgo released the details of more than 50 compact binary merger events over the observations of first run ($\mbox{O1}$), second run ($\mbox{O2}$) and
half of the third run ($\mbox{O3a}$). These observations are grouped 
in the gravitational-wave transient catalog GWTC-1 \cite{GWTC_1}, GWTC-2 \cite{GWTC_2} and GWTC-2.1~\cite{LIGOScientific:2021usb}. With the completion of the second part of the third observing run ($\mbox{O3b}$) in the 
collaboration with LIGO, Virgo and KAGRA (LVK), the total number of gravitational-wave  merger events that includes BBHs, neutron star-black holes (NSBHs) and binary neutron stars (BNS)  has been increased to 90. GWTC-3 is the cumulative catalog
describing all the gravitational-wave transients found in observing runs upto the end of the third observing run (O3) of LIGO-Virgo detectors\cite{GWTC_3}. These BBH detections  enable  determining
the astrophysical properties of the BBH population, such as  mass  and spin distribution of BH populations, the merger rate of such systems and,  their evolution across cosmic time~\cite{LIGOScientific:2018jsj, LIGOScientific:2020kqk, LIGOScientific:2021psn}. Compact binary sources of gravitational wave transient catalog are also used to get some information about the overall sky distribution of events~\cite{Payne2020,Stiskalek2020}. The sky positions of the GW events are localized by performing coherent analyses of multiple detectors' data using  full Bayesian parameter estimation method~\cite{Veitch:2014wba,2019ApJS..241...27A, 2020MNRAS.499.3295R} or rapid Bayesian reconstruction method~\cite{Singer:2015ema}.  During $\mbox{O1}$ and large part of $\mbox{O2}$ only the two LIGO detectors (HL) network  was operational, causing  sky localization areas of the sources to be broader in the range of hundreds to thousands square degrees~\cite{GWTC_1,LIGOScientific:2016dsl}. The addition of  Advanced Virgo to the gravitational-wave  detectors network during  O2 and  O3 has yielded many well-localized  sources observed by all three detectors of the Advanced LIGO, Virgo (HLV) network with sky areas lesser than 100 square degrees.  Such gravitational-wave sources with substantially small areas can be utilized to study the isotropy of the Universe \cite{Payne2020,Stiskalek2020,Mastrogiovanni2022,Reed2022,Cavagli__2020}.  The sky localization areas of sources will improve  further  during the  O4 observation of the  Advanced LIGO, Virgo, and KAGRA (HLVK) network\cite{KAGRA:2013rdx}. With a future five detectors network including the third LIGO detector in India, a significant fraction of gravitational-wave sources will be localized to few square degrees~\cite{Saleem:2021iwi}. 

In this paper, we determine the dipole anisotropy in the extracted black hole mass distribution arising due to our local motion. We also use LIGO/Virgo data \cite{GWOSC} to explore the dipole anisotropy in the distribution of BBH mergers. Such an analysis requires the distribution of number counts per unit mass, which we extract directly from data. We assume a functional form for the distribution law for
the total mass of BBH and fit the parameters of the distribution using the data. We determine the expected dipole anisotropy in this mass distribution of BBH caused by our local
motion. We also determine the dipole anisotropy in mass intensity and mass per source. Next, we extract sky locations of sources from available  posterior distributions of  
BBH merger events  and test the anisotropy in distribution of sources with respect to the CMB direction. 

The  manuscript is structured as follows. In Sec. \ref{sec_2}, we outline the effect of local motion on the observed GW source mass distribution, assuming the power law distribution. We 
present a generalized mass distribution in Sec. \ref{sec_3} and obtain the fit parameters from the GW data for the various cuts on BBH total mass. We also determine the expected dipole
anisotropy in GW sources. In Sec. \ref{sec_4}, we investigate the anisotropy in observed BBH event and then we conclude our results in Sec. \ref{sec_5}.

\section{The Effect Of Local Motion on GW Source Distribution} \label{sec_2}
We start with the assumption that the GW sources are distributed isotropically in the cosmic frame of rest.
 Our local motion with respect to this frame will introduce
two effects, namely, Doppler boosting and wave aberration. The combined effect is expected to change the observed source count and mass intensity (mass weighted number counts) 
 of GW sources in the sky as a function of the sky position. 

Consider a binary system of component masses $m_1$ and $m_2$ in the comoving
frame. The total mass M, chirp mass $\mathcal M_c$ and symmetric mass ratio $\eta$ are defined as $M=m_1+m_2$, $\mathcal M_c= M \eta ^{3/5}$ and $\eta=m_1 m_2/M^2$
respectively. From the observed GW waveform, we can measure the Chirp mass of binary \cite{abbott2016},
\begin{align}
 \mathcal M_c= \frac{c^3}{G} \left[ \frac{5}{96} \,\pi^{-8/3}\, \nu_{obs}^{-11/3} \,\dot\nu_{obs}\right]^{3/5}.
\end{align}
Here, $\mathcal M_c$ is the chirp mass observed in the moving local frame and $\nu_{obs}$ is the GW frequency in the local frame. 

We next determine the anisotropy in the distribution of $\mathcal M_c$ generated due to our motion with respect to the cosmic frame of rest. The chirp mass is affected by the relative motion of the detector with respect to the source. However, this effect will maintain the isotropy of this mass distribution in the cosmic frame of rest and the anisotropy will be generated purely due to our motion relative to this frame. Furthermore, the chirp mass $\mathcal M_c$ and the total mass $M$ are related by $\mathcal M_c = M\eta^{3/5}$, where   $\eta$ depends on the mass ratio $m_1/m_2$. Assuming that the mass ratio is  distributed isotropically in the cosmic frame of rest, our analysis is also directly applicable to the total mass $M$.  

Due to the Doppler effect, the shift in frequency of GW is given 
by \cite{ellis} 
\begin{align}
\nu_{obs}=\nu_{rest} \delta
\end{align}
where, $\delta \approx (1+ \frac{v}{c} \cos(\theta))$ is the Doppler factor, $v$ is the velocity of observer making an angle $\theta$ with the source.
As the observed frequency is Doppler shifted, the mass in the rest frame and the
moving frame is related by
$$(\nu_{obs}^{-11/3} \,\dot\nu_{obs})^{3/5}=\frac{1}{\delta}\,(\nu_{rest}^{-11/3} \,\dot\nu_{rest})^{3/5} $$
This leads to,
\begin{equation}
 \mathcal M_{obs}=\frac{1}{\delta}\mathcal M_{rest},
\label{mshift}
\end{equation}
where $\mathcal M_{obs}$ and $\mathcal M_{rest}$ are chirp masses in the local frame and the cosmic rest frame, respectively. Furthermore, the aberration effect changes the solid angle in the direction of motion, i.e.,  $d\Omega_{obs}=d\Omega_{rest}\delta^{-2}$. 

Let us first assume that the number count distribution has a power law BBH total mass $M$ dependence in the cosmic frame of rest, i.e.,
\begin{align}
 \frac{dN}{d\Omega}(M>M_{min})=k M^{-\alpha}.
\end{align}
$M$ will also follow the relation, Eq.(\ref{mshift}), between rest frame and moving frame.
Let us denote the differential number count per unit solid angle per unit mass by $n(\theta,\phi,M)$, where $(\theta,\phi)$ are polar angles corresponding to the direction of observation or sky location of the source. Assuming isotropy, we have, in the rest frame
\begin{align}
n_{rest}(\theta,\phi, M_{rest}  )\equiv \frac{d^2\,N_{rest}}{d\Omega_{rest}\,  dM_{rest}  }=k\alpha  M_{rest}^{-\alpha - 1}.
\end{align}
Let $d^2\,N_{obs}$ be the number of sources in bin $d\Omega_{obs}\,dM_{obs}$ and $d^2\,N_{rest}$ be the number of sources in the corresponding bin in the rest frame. We have
\begin{align}
d^2\,N_{obs}&=d^2\,N_{rest}=n_{rest}d\Omega_{rest}\,dM_{rest}\nonumber\\ &=k\alpha M_{rest}^{-\alpha-1}\delta^2 d\Omega_{obs}dM_{rest}.
\end{align}
Using the relation for $M_{rest}$, we obtain,
\begin{align}
 d^2N_{obs}=k\alpha\,\delta^{2-\alpha}M_{obs}^{-\alpha-1}d\Omega_{obs}\,dM_{obs}. \label{relpow}
\end{align}
Integrating over $M$ from $M_{min}$ to $\infty$, we get
\begin{align}
\frac{dN_{obs}}{d\Omega_{obs}}=k(M_{min})^{-\alpha}\delta^{2-\alpha}=\frac{dN_{rest}}{d\Omega_{rest}} \delta^{2-\alpha}.
 \end{align}
Hence, the Doppler boosting and aberration, at leading order, will produce a dipole anisotropy in GW source count, given by
\begin{align}
\vec D_N(v)=[2-\alpha](\vec v/c) .
\end{align}

\section{Generalized Mass Distribution} \label{sec_3}

\begin{figure}
\begin{center}
\scalebox{0.8}{\includegraphics*[angle=0,width=\textwidth,clip]{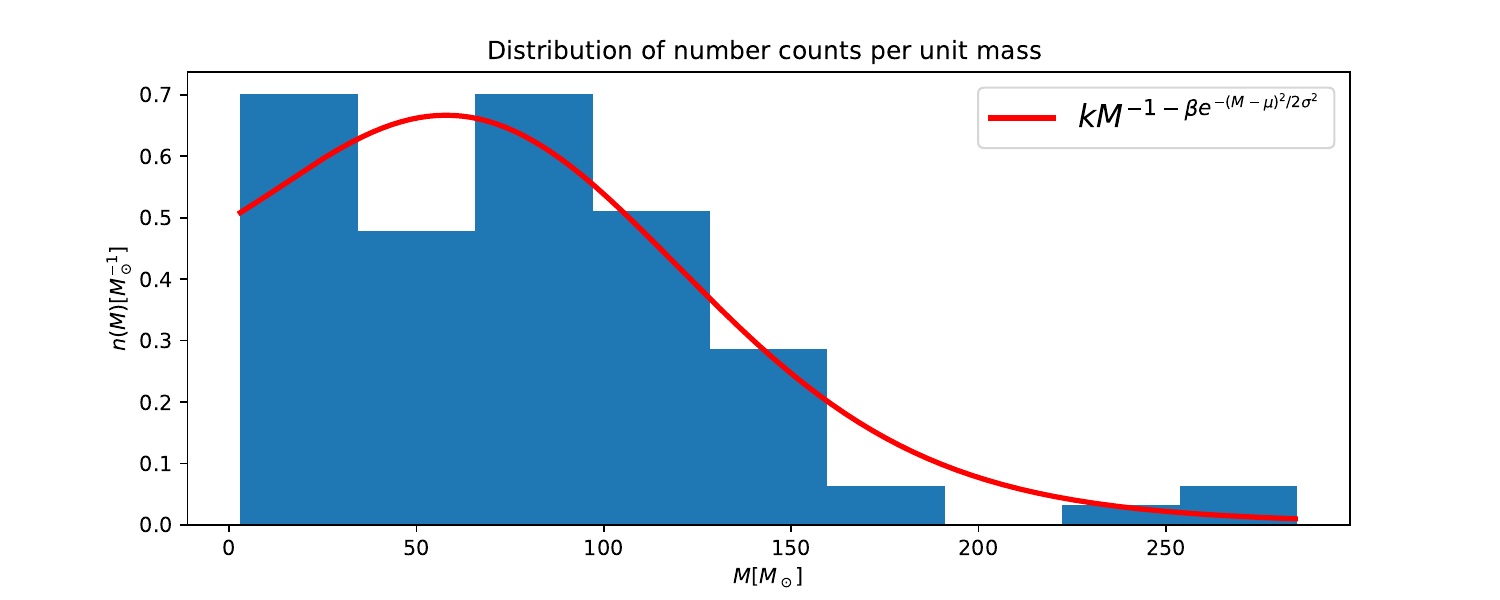}} 
\caption{The fit to the distribution of number counts per unit mass, dN/dM, for the functional form given in Eq.(\ref{ndist}), over the mass range $M>3M_\odot$. The fit parameters
are  $\beta=-1.072,\,\mu=51.695 M_\odot, \, \sigma= -147.71 M_\odot$. 
}
\label{nfit}
\end{center}
\end{figure}

Let us generalize the distribution of total mass M of the binary which takes into account the deviation from a pure power law behavior. We assume the following functional form of $n(\theta,\phi,M_{rest})$
\begin{equation}
n(\theta,\phi,M_{rest})\equiv \frac{d^2N_{rest}} {d\Omega_{rest} dM_{rest}}=k M_{rest}^{-1-f(M_{rest})},
\label{ndist}
\end{equation}
where,
\begin{equation*}
f(M_{rest})=\beta \exp\left[{-\frac{(M_{rest}-\mu)^2}{2\sigma^2}}\right].
\end{equation*}
We fit this functional form to the binary mass data, as shown in Fig.\ref{nfit}. Here we have used the total masses of binary in the detector frame. We find that our functional form of mass distribution is consistent with the other mass models \cite{Talbot2018ApJ,Farrow_2019, Fishbach_2020}.

Following the steps described in Sec. \ref{sec_2}, we obtain the generalized form of Eq. (\ref{relpow}), at leading order in $(v/c) \cos\theta$,
\begin{align}
d^2N_{obs}=k M_{obs}^{-1-f(M_{obs})}\left[1-M_{obs} \,\frac{v}{c} \cos \theta \,\frac{df(M_{obs})}{dM}\, \ln (M_{obs})\right ]\delta^{2-f(M_{obs})} d \Omega_{obs} dM_{obs}.
\label{d2n}
\end{align}
Integrating $M_{obs}$ from $M_{min}$ to $M_{max}$, we get  
\begin{align}
\frac{dN_{obs}}{d\Omega_{obs}}&=k\left(I_1+ (2I_1+I_2) \frac{v}{c}\,cos \theta \right)\nonumber\\
&= \frac{dN_{rest}}{d\Omega_{rest}} \left(1+a \frac{v}{c} cos \theta \right),
\label{n_obs}
\end{align}
where,

\begin{align}
&\frac{dN_{rest}}{d\Omega_{rest}}= kI_1\\
 &a=2+ \frac{I_2}{I_1}\\
 &I_1= \int_{M_{min}}^{M_{max}}  M_{obs}^{-1-f(M_{obs})} dM_{obs}\\
 &I_2= -\int_{M_{min}}^{M_{max}}[f(M_{obs})+M_{obs} f'(M_{obs})\ln (M_{obs}) \, M_{obs}^{-1-f(M_{obs})}dM_{obs}.
\end{align}
If the sources are distributed isotropically in the cosmic rest frame,  the observed amplitude of Dipole anisotropy in source counts will be 
\begin{align}
|D_N|=a \frac{v}{c} \cos\theta.
\end{align}
Integrating $I_1$ and $I_2$ from $M_{min} =3 M_{\odot}$ to $M_{max}=400 M_\odot$ and using the fit parameters, we obtain 
\begin{align}
I_1=179.3, \quad I_2=-1.56, \quad a=1.991
\end{align}
Taking the local velocity to be, v=370 Km/s, we expect that the amplitude of dipole anisotropy will be $ |D_N| \approx 2.46 \times 10^{-3}$ for $M> 3M_\odot$.\\

We next generalise this calculation to mass intensity, $M_I$. This is
defined as the number counts weighted by mass, i.e.,
\begin{align}
d^2M_I=M d^2 N.
\end{align}
We have,
\begin{align}
d^2M_{I,obs}=M_{obs} \,d^2N_{obs}=\,M_{obs}n_{rest} d\Omega_{rest} dM_{rest}.
\end{align}
Using Eq.(\ref{d2n}) and integrating over $M_{obs}$ from $M_{min}$ to $M_{max}$, we obtain 
\begin{align}
\frac{dM^{}_{I,obs}}{d\Omega_{obs}}=\frac{dM^{}_{I,rest}}{d\Omega_{rest}} \left(1+b \frac{v}{c} \cos \theta \right)
\label{M_obs},
\end{align}

where,
\begin{align}
&\frac{dM^{}_{I,rest}}{d\Omega_{rest}}= kI_3\\
 &b=2+ \frac{I_4}{I_3}\\
 &I_3= \int_{M_{min}}^{M_{max}}  M_{obs}^{-f(M_{obs})} dM_{obs}\\
 &I_4= -\int_{M_{min}}^{M_{max}} [f(M_{obs})+M_{obs} f'(M_{obs}) \ln M_{obs}]\, M_{obs}^{-f(M_{obs})}dM_{obs}
\end{align}
Therefore, the observed dipole anisotropy in mass intensity is given by
\begin{align}
D_M=b \frac{v}{c}\cos \theta.
\end{align}

We  also consider another observable, mass per source in angular bin of size $d\Omega$, defined as,
\begin{align}
\frac{dm}{dN}= \frac{(dM_I/d \Omega)}{(dN/d\Omega)}.
\end{align}
Using Eq.(\ref{n_obs}) and Eq.(\ref{M_obs}), the observed value of mass per source, at leading order in $(v/c) \cos(\theta)$, is given by 
\begin{align}
\frac{dm^{}_{obs}}{dN_{obs}}=\frac{dm^{}_{rest}}{dN_{rest}} \left(1-d \frac{v}{c} \cos \theta\right),
\end{align}
where $d= (a-b)$.
Hence, the observed dipole anisotropy in mass per source will be 
\begin{align}
D_m=-d \frac{v}{c} \cos\theta.
\end{align}
It is quite interesting that the dipole in this parameter is opposite to that in number counts. Hence it may be very useful to test the kinematic origin of the dipole.

\begin{table}[h!]
\centering
\begin{tabular}{ p{1.5cm} p{1.5cm} p{1.5cm} p{1.5cm} p{1.5cm}  }
 \hline
 $M(M_\odot)>$ & 3 & 10 & 20 & 30\\
 \hline
 $\beta$   & -1.072 &-1.074 &-1.106 & -1.082\\
 $\mu$      & 51.695 & 60.531 & 27.771 & 87.903\\
 $\sigma$   & -147.71 & -134.69 & -167.52 & -78.874\\
 a          & 1.991 & 1.95 & 1.846 & 1.879\\
 b          & 1.039 & 1.028 & 1.01 & 0.991\\
 d          & 0.952 & 0.922 & 0.836 & 0.887\\
 \hline
\end{tabular}
\caption{The fit parameters corresponding to Eq.(\ref{ndist}) for various cuts on mass and values of $a$, $b$ and $d$ for each set of parameters are given.}
\label{tbl1}
\end{table}

The fit parameters corresponding to Eq.(\ref{ndist}) and values of $a$, $b$ and $d$ are given in Table \ref{tbl1} for different cuts on the mass of binary. We point out that the parameter $d$ is also quite large. This is in contrast to a pure power law, in which case we do not expect any dipole anisotropy in mass per source in analogy to brightness per source in the case of radio data \cite{tiwari2015}.
In comparing with the results of matter dipole \cite{ellis} we find that the signal in number counts is reduced by a factor of 2. Hence for a three sigma detection we would require four times as many sources. Using the calculation in \cite{ellis} we find that this number comes out to be approximately $8\times 10^5$. This may be within reach of future next generation detectors, such as Einstein Telescope and Cosmic Explorer \cite{Reitze2019,Punturo_2010,Dwyer2015}. These detectors might observe $\sim 10^5$ GWs per year coming from BBHs and BNSs mergers with signal to noise ratio threshold more than 9, as discussed in \cite{Mastrogiovanni2022}. This number of GWs events are estimated for redshift shell of  $z=2$. Therefore, it may take $\sim$ 8(2) years of observation to detect the $ 3 \sigma$($2\sigma$) level dipole anisotropy in source number counts. In any case, given that this effect will also be observed in galaxy surveys, even a one sigma detection, requiring an order of magnitude fewer sources, would be very useful to establish consistency. We next determine the number of sources needed to observe the dipole in mass per source. This is dependent on the standard deviation in the mass distribution. Given that the velocity is about 370 Km/sec and $d=0.95$ we expect the dipole amplitude to be approximately $10^{-3}$. We find that a three sigma detection in this case requires about 4 million sources. It may take $\sim$ 40(7) years of observation to detect the $3 \sigma$($2\sigma$) level dipole anisotropy in mass per source.

\section{Data Analysis} \label{sec_4}
The currently available GW data is rather limited and we do not expect to obtain a signal of local velocity in this data. The resulting dipole anisotropy signal is of similar strength as in radio sources and hence we require roughly $8\times 10^5$ sources to extract a signal at 3 sigma significance \cite{ellis}. However, the required signal is within the reach of gravitational wave detectors in near future. In order to illustrate the procedure we extract the dipole in the mass per source. 

The posterior samples containing information about the parameters of events were taken from the LIGO/Virgo data released via the Gravitational-Wave Open Science
Center (GWOSC)\cite{GWOSC}. We extracted the mass and location of each source using the given probability distribution. For few sources the probability distribution of their sky location (RA, DEC) are uni-modal and are completely in either forward or backward hemisphere with respect to the CMB dipole direction ($RA=167.9^o$, $DEC=-6.93^o$). Other sources have multi-modal probability distribution of sky location and spread across both the hemispheres. As CMB dipole is near to the Celestial equator, so we assume that the position of any event with respect to the CMB dipole direction will only be decided by the angle RA of that event. To estimate the probability of occurrence of these sources in any one of the hemispheres we use the idea of finite mixture modelling for the probability distribution of RA.

\begin{figure}[ht!]
\begin{center}
\scalebox{0.8}{\includegraphics[angle=0,width=\textwidth]{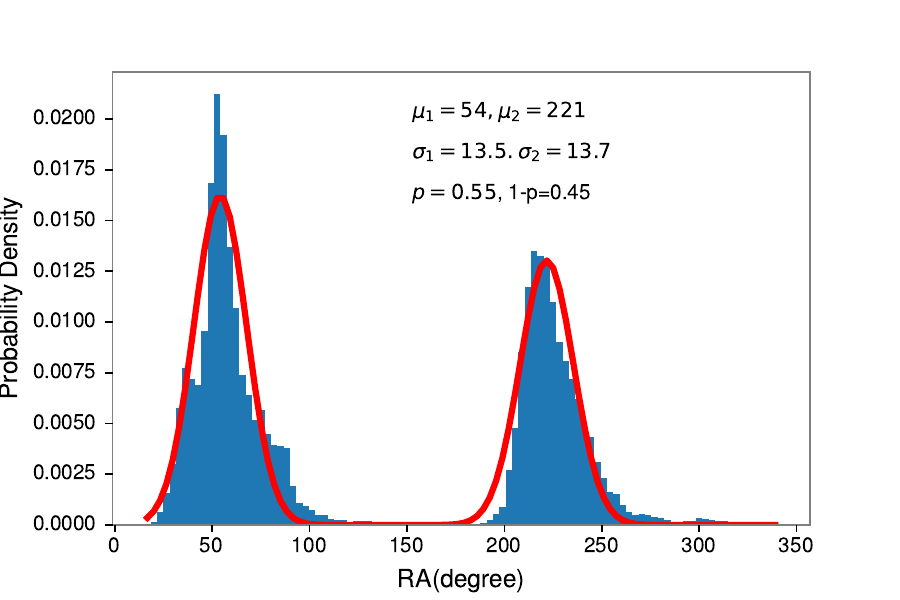}} 
\caption{Bimodal probability distribution of RA and approximated Gaussian distributions.}
\label{bimodal}
\end{center}
\end{figure}

Let f(x) be the probability distribution of RA, we can write it as
\begin{align}
 f(x)=p f_1(x)+(1-p) f_2(x)
\end{align}
 where we have taken $f_1$ and $f_2$ both Gaussian and there would be 5 parameters to estimate from the data, $p,\sigma_1,\sigma_2,\mu_1$, and $\mu_2$, defined as mixture proportion, the standard deviations and means respectively for the distributions $f_1$ and $f_2$. One such fitting is shown in Fig. [\ref{bimodal}] for the event 
 GW$191219\_163120$.It may be possible to improve the fit using a different functional form. This may improve the extracted position coordinates of the sources and hence lead to better determination of the dipole. However, the dominant source of error is the intrinsic fluctuations in number counts and the error in mass values or positions is comparatively negligible. Hence, we find that our analysis is sufficiently reliable and postpone further refinements to future research.
 
\begin{figure}[ht!]
\begin{center}
\scalebox{0.7}{\includegraphics[angle=0,width=\textwidth]{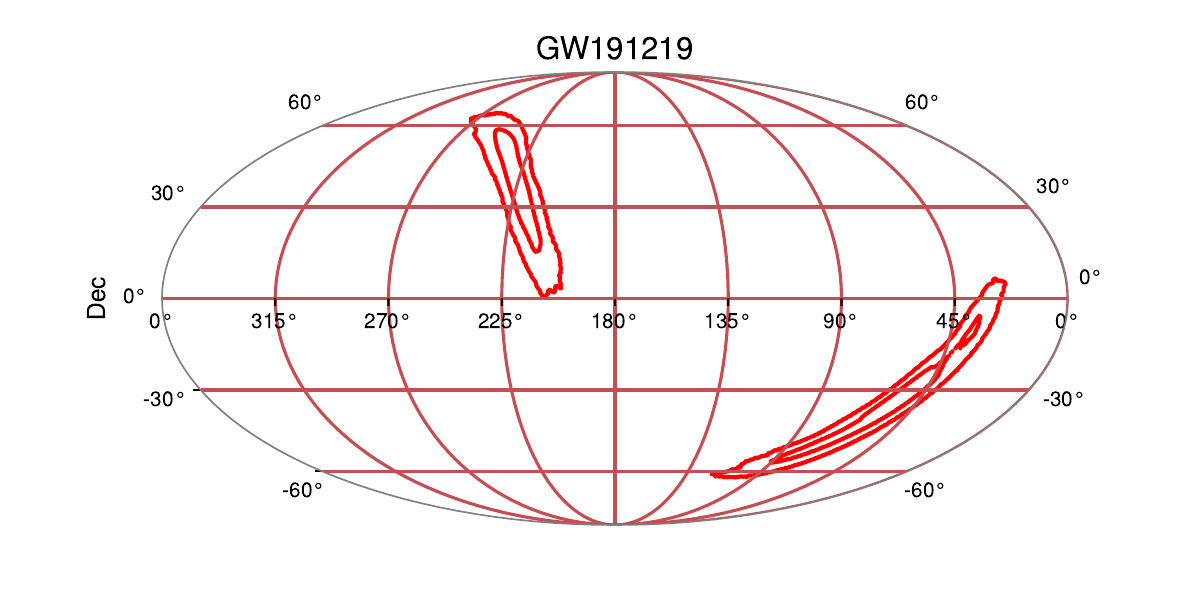}}
\caption{Skymap of the event GW$191219\_163120$}
\label{GW191219}
\end{center}
\end{figure}

From Fig. [\ref{bimodal}], we can say that probability of the event at $RA_1=54$ is 0.55 and probability of event at $RA_2=221$ is 0.45.
To get an estimate of the declination angle (DEC) of the event at $RA_1$ and $RA_2$, we look into the skymap of the given event and approximated the corresponding angles as $DEC_1=-35$ and $DEC_2=35$ directly from the given skymap, as shown in Fig. [\ref{GW191219}]

\begin{figure}[ht!]
\begin{center}
\scalebox{0.8}{\includegraphics[angle=0,width=\textwidth]{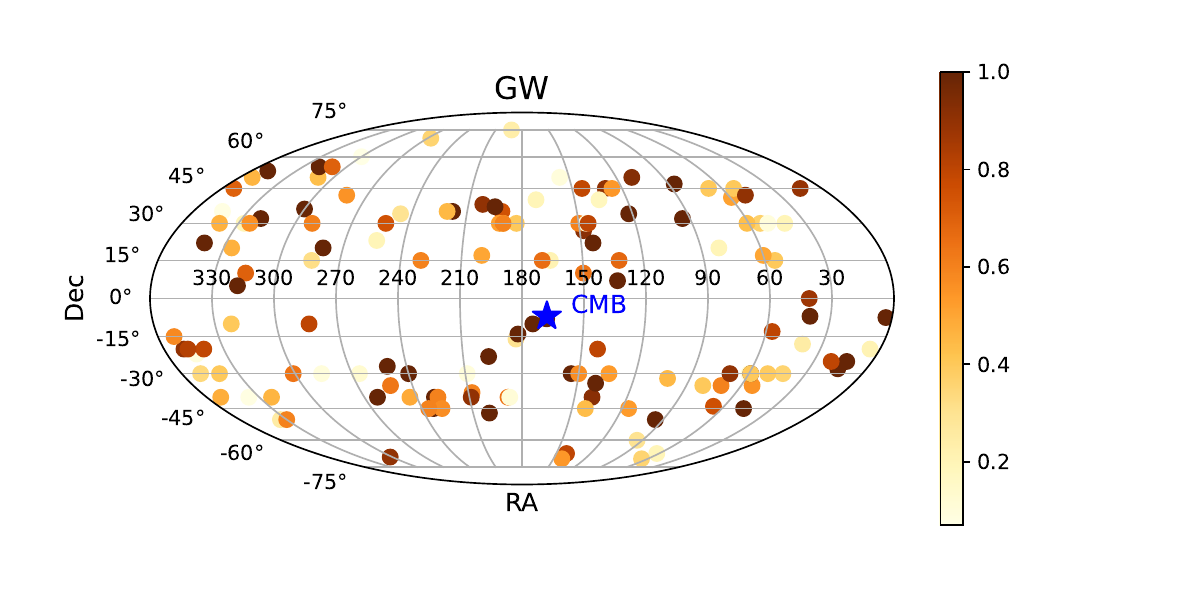}} 
\caption{Sky location of all GW events with the colorbar showing their probability at that position.}
\label{BBH_loc}
\end{center}
\end{figure}

We have done the same analysis for all the events and determine the angles RA, DEC and corresponding probabilities of each event. The resulting skymap of all events weighted with their probability is shown in Fig.[\ref{BBH_loc}]. Here colorbar represents the probability of the event. 

We determine the cosine of angle ($\cos\theta$) between unit vector along the CMB dipole and unit vector along the direction of each source. In forward hemisphere with respect to CMB dipole, $\cos\theta$ will be positive and in backward hemisphere $\cos\theta$ will be negative.
We use the masses of events in detector frame. The results are shown in Table \ref{GW_data}\ .

In order to determine if the sources are distributed isotropically we determine the observable mean value of the mass per source in each hemisphere. We define mean mass per source $m$ as: 
\begin{align}
   m=& \dfrac{\sum\limits_{i=1}^N M_i p_i }{\sum\limits_{i=1}^N  p_i}\,.
\end{align}
where $M_i$ is the observed mass of the source and $p_i$ the probability that the source lies in a particular hemisphere.
From data we find that the mean mass per source in forward hemisphere is $86.19 ^{5.64}_{2.55}$ $M_\odot$ whereas in backward hemisphere it is $84.6 ^{8.04}_{3.0}$ $M_\odot$. We find that the difference between the two is not statistically significant and hence the data is consistent with isotropy.

We perform a more detailed analysis by determining the best fit value of the dipole amplitude. We test the following model for the mass distribution on sky,
\begin{equation}
    M_{\rm th}(\theta,\phi) = M_0 + \delta M\cos\theta 
\end{equation}
where $\theta$ is the polar angle in a frame with $z$-axis pointing along the CMB dipole axis and $\phi$ being the corresponding azimuthal angle. Here $M_0$ is the monopole term and $\delta M$ the dipole amplitude. We determine the two parameters $M_0$ and $\delta M$ by minimizing the weighted least square difference. This is defined as
\begin{equation}
    \Delta = \frac{\sum\limits_{i=1}^N p_i\left(M_i - M_0 -\delta M\cos\theta  \right)^2 } {\sum\limits_{i=1}^N p_i}
\end{equation}
 The best fit value of parameters is found to be: $M_0= 85.3$ $M_\odot$ and $\delta M=6.2$ $M_\odot$. 
We point out that sample has dispersion which is much larger than the error bars on individual masses. Hence we ignore the errors in masses while determining the best fit values.
We determine the statistical significance of the dipole by simulations. We generate 100 random samples by randomly permuting the mass values among different sources while keeping their angular positions fixed. We determine the best fit values of each of these samples and find that among these 46 lead to dipole values larger than that of the data sample. Hence we conclude that the data is consistent with isotropy.

\section{Conclusions} \label{sec_5}
The search for dipole anisotropy in large scale structures is turning out to be very interesting due to results obtained by radio \cite{Singal:2011,Gibelyou:2012,Rubart:2013,tiwari2015} and infrared surveys \cite{Secrest:2020}. Both show a significant deviation from the amplitude expected on the basis of the CMB dipole while the direction is found to be in reasonable agreement. Such a trend has raised the possibility that the Universe might be intrinsically anisotropic and has lead to considerable theoretical effort \citep{Perivolaropoulos:2021}. The GW observations open up a new and independent domain to test this phenomenon. Here we have determined the signal of dipole anisotropy in the GW observations predicted due to our velocity with respect to the cosmic frame of rest. We considered three different observables, the number counts, the number counts weighted by mass (also called mass intensity) and the mean mass per source. We assume a functional form for the mass dependence of the number count distribution and extract the parameters of this function directly from data. We find that all three observables acquire a significant dipole anisotropy due to our local motion, which can be tested reliably once a sufficiently large sample of GW becomes available.

In order to illustrate our procedure, we have also analysed the available data in order to determine whether the mean mass per source shows any deviation in the two opposite hemispheres in the direction of the CMB dipole. We find that the sample is consistent with isotropy. The data sample is currently very small and considerably more data is required to determine the signal of local motion. The signal is within the reach of future gravitational wave observatories and we look forward to a more detailed study of this phenomenon with future data.

\bigskip
\acknowledgments{We thank Marco Finetti for useful discussions. GK acknowledges the Department of Applied Physics, Gautam Buddha University, Greater Noida for initial support. Naveen K Singh acknowledges the School of Physics and Astronomy, Sun Yat-sen University, Zhuhai Campus, China  for the financial support during the initial period of the project. This research has made use of data, software and/or web tools obtained from the Gravitational Wave Open Science Center (https://www.gw-openscience.org/ ), a service of LIGO Laboratory, the LIGO Scientific Collaboration and the Virgo Collaboration. LIGO Laboratory and Advanced LIGO are funded by the United States National Science Foundation (NSF) as well as the Science and Technology Facilities Council (STFC) of the United Kingdom, the Max-Planck-Society (MPS), and the State of Niedersachsen/Germany for support of the construction of Advanced LIGO and construction and operation of the GEO600 detector. Additional support for Advanced LIGO was provided by the Australian Research Council. Virgo is funded, through the European Gravitational Observatory (EGO), by the French Centre National de Recherche Scientifique (CNRS), the Italian Istituto Nazionale della Fisica Nucleare (INFN) and the Dutch Nikhef, with contributions by institutions from Belgium, Germany, Greece, Hungary, Ireland, Japan, Monaco, Poland, Portugal, Spain.}

\bigskip
\LTcapwidth=0.8\textwidth
\begingroup
\renewcommand\arraystretch{1}
\begin{longtable}{ l p{3cm} p{2cm} p{2cm} p{2cm} p{1cm} p{2cm} c}
\caption{ \large The masses of GW events in detector frame, extracted location of GW sources, probability(p) of occurance of event in forward hemishpere ($\cos \theta >0$) or in backward hemisphere ($\cos \theta <0$), and direction with respect to CMB dipole. }\\
\label{GW_data}\\
\hline
\\
\scriptsize
\textbf{S.No} &\textbf{Event} &\textbf{M$_{det}(M_\odot)$} & \textbf{RA(deg)} &\textbf{DEC(deg)} &\textbf{p} &\textbf{$\cos\theta$} 
\\
\\
\hline
\\

1 &GW150914\_095045 &$70.92_{-3.57}^4$ &141 &-67 &1 &0.45697237 \\
2 &GW151012\_095443 &$46.4_{-4.5}^{12.4}$ &59.3 &41 &0.51 &-0.31812088 \\
3 &GW151012\_095443 &$46.4_{-4.5}^{12.4}$ &244 &-40 &0.49 &0.26023751 \\
4 &GW151226\_033853 &$23.7_{-1.4}^{9.2}$ &53 &45 &0.4 &-0.38085937 \\
5 &GW151226\_033853 &$23.7_{-1.4}^{9.2}$ &208 &-38 &0.6 &0.6726463 \\
6 &GW170104\_101158 &$60.4_{-4}^4$ &345 &-22 &0.1 &-0.87403264 \\
7 &GW170104\_101158 &$60.4_{-4}^4$ &130 &45 &0.9 &0.46857325 \\
8 &GW170608\_020116 &$19.8_{-0.33}^{2.28}$ &122 &34 &1 &0.50525259 \\
9 &GW170729\_185629 &$123_{-21}^{21}$ &302 &-69 &0.9 &-0.13492819 \\
10 &GW170729\_185629 &$123_{-21}^{21}$ &156 &50 &0.1 &0.53195002 \\
11 &GW170809\_082821 &$70.6_{-4.5}^6$ &15 &-28 &1 &-0.72362404 \\
12 &GW170814\_103043 &$68.8_{-3}^3$ &47 &-45 &1 &-0.27515847 \\
13 &GW170817 &$2.75_{-0.016}^1$ &197 &-23 &1 &0.84557982 \\
14 &GW170818\_022509 &$75.76_6^6$ &341 &22 &1 &-0.95894268 \\
15 &GW170823\_131358 &$90.8_{-9}^{11}$ &54 &30 &0.37 &-0.40862792 \\
16 &GW170823\_131358 &$90.8_{-9}^{11}$ &252 &-35 &0.63 &0.15279336 \\
17 &GW190403\_051519 &$236.18_{-54}^{40}$ &334 &-50 &0.25 &-0.52697769 \\
18 &GW190403\_051519 &$236.18_{-54}^{40}$ &191 &35 &0.75 &0.67876321 \\
19 &GW190408\_181802 &$55.6_{-3.83}^{3.42}$ &349 &53 &1 &-0.69366892 \\
20 &GW190412 &$42.1_{-4.51}^{5.36}$ &218 &35 &1 &0.45240024 \\
21 &GW190413\_052954 &$90.84_{-13}^{15}$ &66.3 &-44 &0.76 &-0.05977149 \\
22 &GW190413\_052954 &$90.84_{-13}^{15}$ &192 &75 &0.24 &0.11798748 \\
23 &GW190413\_134308 &$135_{-17.75}^{17.73}$ &154 &-30 &1 &0.8948518 \\
24 &GW190421\_213856 &$108.9_{-12.44}^{15.31}$ &200 &-47 &1 &0.66175767 \\
25 &GW190426\_190642 &$312.16_{-44}^{61}$ &57 &-13 &0.8 &-0.31791357 \\
26 &GW190426\_190642 &$312.16_{-44}^{61}$ &254 &23 &0.2 &0.01500671 \\
27 &GW190503\_185404 &$91.63_{-12.13}^{11.26}$ &95 &-50 &1 &0.280053 \\
28 &GW190512\_180714 &$45.31_{-2.8}^{3.88}$ &250 &-27 &1 &0.17634627 \\
29 &GW190513\_205428 &$73.6_{-6.72}^{12.57}$ &50 &42 &0.9 &-0.42593418 \\
30 &GW190513\_205428 &$73.6_{-6.72}^{12.57}$ &286 &-30 &0.1 &-0.34459989 \\
31 &GW190514\_065416 &$114.64_{-19}^{20}$ &60 &-30 &0.1 &-0.2039057 \\
32 &GW190514\_065416 &$114.64_{-19}^{20}$ &202 &38 &0.9 &0.57346969 \\
33 &GW190517 055101 &$85.4_{-7.31}^{9.7}$ &233 &-45 &1 &0.38085937 \\
34 &GW190519\_153544 &$154.84_{-18}^{18}$ &353 &45 &0.7 &-0.78447906 \\
35 &GW190519\_153544 &$154.84_{-18}^{18}$ &183 &-16 &0.3 &0.95454912 \\
36 &GW190521\_030229 &$243.3_{-36}^{58}$ &352 &-40 &0.48 &-0.68094522 \\
37 &GW190521\_030229 &$243.3_{-36}^{58}$ &192 &30 &0.52 &0.72443384 \\
38 &GW190521\_074359 &$92.36_{-5}^6$ &280 &20 &1 &-0.39221934 \\
39 &GW190527\_092055 &$84.43_{-10.49}^{54.92}$ &301 &-30 &0.64 &-0.52708111 \\
40 &GW190527\_092055 &$84.43_{-10.49}^{54.92}$ &42 &-30 &0.36 &-0.44377511 \\
41 &GW190602\_175927 &$171.9_{-20.58}^{22.64}$ &71 &-35 &0.6 &-0.02848558 \\
42 &GW190602\_175927 &$171.9_{-20.58}^{22.64}$ &81 &-35 &0.4 &0.11318094 \\
43 &GW190620\_030421 &$140.5_{-22}^{20}$ &40 &-18 &0.25 &-0.54266686 \\
44 &GW190620\_030421 &$140.5_{-22}^{20}$ &252 &30 &0.75 &0.02804227 \\
45 &GW190630\_185205 &$69.7_{-4}^5$ &340 &-20 &0.79 &-0.88270748 \\
46 &GW190630\_185205 &$69.7_{-4}^5$ &166 &15 &0.21 &0.92711366 \\
47 &GW190701\_203306 &$129.64_{-14.75}^{16.61}$ &40 &-7 &1 &-0.59054774 \\
48 &GW190706\_222641 &$180.3_{-27.8}^{22.4}$ &336 &-40 &0.08 &-0.66654857 \\
49 &GW190706\_222641 &$180.3_{-27.8}^{22.4}$ &148 &27 &0.92 &0.77690516 \\
50 &GW190707\_093326 &$23.36_{-0.69}^{1.54}$ &310 &50 &0.43 &-0.59593626 \\
51 &GW190707\_093326 &$23.36_{0.69}^{1.54}$ &150 &-30 &0.57 &0.87841258 \\
52 &GW190708\_232457 &$37.12_{-1.67}^{2.83}$ &350 &-30 &0.34 &-0.79879279 \\
53 &GW190708\_232457 &$37.12_{-1.67}^{2.83}$ &170 &15 &0.66 &0.92699685 \\
54 &GW190719\_215514 &$91.43_{-15}^{74}$ &340 &-30 &0.4 &-0.79121117 \\
55 &GW190719\_215514 &$91.43_{-15}^{74}$ &150 &30 &0.6 &0.75775595 \\
56 &GW190720\_000836 &$25.3_{-1.52}^{4.28}$ &300 &36 &1 &-0.6093442 \\
57 &GW190725\_174728 &$21.83_{-1.29}^{9.31}$ &284 &-10 &0.8 &-0.40913847 \\
58 &GW190725\_174728 &$21.83_{-1.29}^{9.31}$ &81 &20 &0.2 &0.00917921 \\
59 &GW190727\_060333 &$104.61_{-11}^{13}$ &352 &50 &0.46 &-0.72888692 \\
60 &GW190727\_060333 &$104.61_{-11}^{13}$ &142 &-70 &0.54 &0.4187993 \\
61 &GW190728\_064510 &$24_{-0.73}^{5.27}$ &315 &10 &0.7 &-0.84177514 \\
62 &GW190728\_064510 &$24_{-0.73}^{5.27}$ &94 &-60 &0.3 &0.24213604 \\
63 &GW190731\_140936 &$109.4_{-14.32}^{14.76}$ &66 &-70 &0.36 &0.0433694 \\
64 &GW190731\_140936 &$109.4_{-14.32}^{14.76}$ &188 &-40 &0.64 &0.79168889 \\
65 &GW190803\_022701 &$100_{-11.97}^{14.14}$ &94 &32 &1 &0.16951977 \\
66 &GW190805\_211137 &$147.36_{-21}^{21}$ &350 &-20 &0.9 &-0.89093402 \\
67 &GW190805\_211137 &$147.36_{-21}^{21}$ &146 &30 &0.1 &0.73733111 \\
68 &GW190814 &$27.21_{-1.32}^{1.44}$ &13 &-25 &1 &-0.7637364 \\
69 &GW190828\_063405 &$78.57_{-5}^6$ &327 &30 &0.2 &-0.86346248 \\
70 &GW190828\_063405 &$78.57_{-5}^6$ &142 &-20 &0.8 &0.88039924 \\
71 &GW190828\_065509 &$44_{-4}^5$ &344 &35 &0.08 &-0.8804903 \\
72 &GW190828\_065509 &$44_{-4}^5$ &140 &-40 &0.92 &0.74961435 \\
73 &GW190910\_112807 &$100.83_{-8}^9$ &240 &-30 &1 &0.32456233 \\
74 &GW190915\_235702 &$78.4_{-7.99}^{8.37}$ &195 &37 &1 &0.63314856 \\
75 &GW190916\_200658 &$119.7_{-22}^{39}$ &352 &-15 &0.58 &-0.92518688 \\
76 &GW190916\_200658 &$119.7_{-22}^{39}$ &190 &30 &0.42 &0.73620694 \\
77 &GW190917\_114630 &$13.52_{-3.3}^{3.21}$ &262.5 &-40 &1 &0.01656948 \\
78 &GW190924\_021846 &$15.45_{-0.6}^{3.22}$ &133.56 &7 &1 &0.79885827 \\
79 &GW190925\_232845 &$44_2^{-5.27}$ &182 &-14 &1 &0.96337703 \\
80 &GW190926\_050336 &$94.93_{-13}^{44}$ &340 &30 &0.47 &-0.9118678 \\
81 &GW190926\_050336 &$94.93_{-13}^{44}$ &134 &-30 &0.53 &0.77388862 \\
82 &GW190929\_012149 &$144_{-19}^{31}$ &282 &42 &0.55 &-0.38196681 \\
83 &GW190929\_012149 &$144_{-19}^{31}$ &102 &-32 &0.45 &0.4076923 \\
84 &GW190930\_133541 &$23.26_{-0.97}^{10.45}$ &319 &55 &1 &-0.59731342 \\
85 &GW191103\_012549 &$23.47_{-0.68}^{4.58}$ &267 &70 &0.36 &-0.1670782 \\
86 &GW191103\_012549 &$23.47_{-0.68}^{4.58}$ &150 &10 &0.64 &0.9093393 \\
87 &GW191105\_143521 &$22.38_{-0.5}^{2.35}$ &348 &-20 &0.82 &-0.89155909 \\
88 &GW191105\_143521 &$22.38_{-0.5}^{2.35}$ &136 &40 &0.18 &0.5680422 \\
89 &GW191109\_010717 &$140.43_{-16.71}^{21.32}$ &230 &-40 &1 &0.43339286 \\
90 &GW191113\_071753 &$43.36_{-9.88}^{13.45}$ &55 &15 &0.4 &-0.40434715 \\
91 &GW191113\_071753 &$43.36_{-9.88}^{13.45}$ &228 &-40 &0.6 &0.45663057 \\
92 &GW191126\_115259 &$26.54_{-0.9}^{4.14}$ &323 &-40 &0.46 &-0.61220318 \\
93 &GW191126\_115259 &$26.54_{-0.9}^{4.14}$ &126 &45 &0.54 &0.43714558 \\
94 &GW191129\_134029 &$20.1_{-0.64}^{2.94}$ &330 &-50 &0.6 &-0.51477605 \\
95 &GW191129\_134029 &$20.1_{-0.64}^{2.94}$ &183 &30 &0.4 &0.76968704 \\
96 &GW191204\_110529 &$62.76_{-5.32}^{10.25}$ &284 &15 &0.32 &-0.45307229 \\
97 &GW191204\_110529 &$62.76_{-5.32}^{10.25}$ &132 &15 &0.68 &0.74549564 \\
98 &GW191204\_171526 &$22.74_{-0.48}^{1.94}$ &70 &-30 &0.9 &-0.05783255 \\
99 &GW191204\_171526 &$22.74_{-0.48}^{1.94}$ &209 &-30 &0.1 &0.70816562 \\
100 &GW191215\_223052 &$58.41_{-3.68}^{4.81}$ &326 &20 &0.47 &-0.90677818 \\
101 &GW191215\_223052 &$58.41_{-3.68}^{4.81}$ &116 &-45 &0.53 &0.51843982 \\
102 &GW191216\_213338 &$21.2_{-0.66}^{2.93}$ &320 &32 &1 &-0.80793859 \\
103 &GW191219\_163120 &$36.03_{-2.61}^{2.22}$ &54 &-35 &0.55 &-0.2602422 \\
104 &GW191219\_163120 &$36.03_{-2.61}^{2.22}$ &221 &35 &0.45 &0.41903645 \\
105 &GW191222\_033537 &$119.21_{-13}^{15.78}$ &50 &30 &0.1 &-0.46260687 \\
106 &GW191222\_033537 &$119.21_{-13}^{15.78}$ &209 &-40 &0.9 &0.65060233 \\
107 &GW191230\_180458 &$145.18_{-18.95}^{20.39}$ &59 &-30 &1 &-0.218143 \\
108 &GW200112\_155838 &$79.08_{-5.08}^{6.46}$ &291 &30 &0.61 &-0.52981134 \\
109 &GW200112\_155838 &$79.08_{-5.08}^{6.46}$ &50 &-30 &0.39 &-0.34195024 \\
110 &GW200115\_042309 &$7.83_{-1.78}^{1.85}$ &41 &0 &0.87 &-0.59603374 \\
111 &GW200115\_042309 &$7.83_{-1.78}^{1.85}$ &266 &-30 &0.13 &-0.06080426 \\
112 &GW200128\_022011 &$116.22_{-13.45}^{17.3}$ &61 &30 &0.43 &-0.31024455 \\
113 &GW200128\_022011 &$116.22_{-13.45}^{17.3}$ &228 &-45 &0.57 &0.43522604 \\
114 &GW200129\_065458 &$74.61_{-3.83}^{4.45}$ &318 &5 &1 &-0.86780468 \\
115 &GW200202\_154313 &$19_{-0.34}^{1.99}$ &144 &22 &1 &0.79628983 \\
116 &GW200208\_130117 &$91.43_{-10}^{11.41}$ &140 &-34 &1 &0.79479254 \\
117 &GW200208\_222617 &$101.61_{-39.7}^{168.47}$ &13 &45 &0.89 &-0.72097287 \\
118 &GW200208\_222617 &$101.61_{-39.7}^{168.47}$ &187 &-40 &0.11 &0.79614105 \\
119 &GW200209\_085452 &$98.5_{-15.83}^{19.22}$ &5 &-20 &0.21 &-0.85032302 \\
120 &GW200209\_085452 &$98.5_{-15.83}^{19.22}$ &144 &45 &0.79 &0.5564351 \\
121 &GW200210\_092254 &$32.05_{-4.72}^{7.93}$ &322 &-10 &0.4 &-0.85846762 \\
122 &GW200210\_092254 &$32.05_{-4.72}^{7.93}$ &190 &30 &0.6 &0.73620694 \\
123 &GW200216\_220804 &$135.12_{-32.22}^{30.17}$ &310 &55 &0.7 &-0.5481296 \\
124 &GW200216\_220804 &$135.12_{-32.22}^{30.17}$ &246 &34 &0.3 &0.10223177 \\
125 &GW200219\_094415 &$102.51_{-11.92}^{14.11}$ &21 &-25 &0.8 &-0.70269259 \\
126 &GW200219\_094415 &$102.51_{-11.92}^{14.11}$ &172 &40 &0.2 &0.68094522 \\
127 &GW200220\_061928 &$284.5_{-42.91}^{62.86}$ &168 &-8 &1 &0.99982413 \\
128 &GW200220\_124850 &$110.86_{-14}^{16}$ &59 &-30 &0.4 &-0.218143 \\
129 &GW200220\_124850 &$110.86_{-14}^{16}$ &230 &15 &0.6 &0.41745519 \\
130 &GW200224\_222234 &$94.86_{-7.18}^{8.27}$ &174.7 &-10 &1 &0.99168784 \\
131 &GW200225\_060421 &$41.23_{-4.01}^{2.99}$ &300 &60 &0.07 &-0.43725605 \\
132 &GW200225\_060421 &$41.23_{-4.01}^{2.99}$ &110 &50 &0.93 &0.24665263 \\
133 &GW200302\_015811 &$73.7_{-7.94}^{14.84}$ &68 &45 &0.4 &-0.20600118 \\
134 &GW200302\_015811 &$73.7_{-7.94}^{14.84}$ &236 &-45 &0.6 &0.34713249 \\
135 &GW200306\_093714 &$62_{-12.19}^{14.11}$ &41 &30 &0.2 &-0.57650867 \\
136 &GW200306\_093714 &$62_{-12.19}^{14.11}$ &145 &30 &0.8 &0.73161338 \\
137 &GW200308\_173609 &$214.68_{-130}^{360}$ &324 &30 &0.56 &-0.84631105 \\
138 &GW200308\_173609 &$214.68_{-130}^{360}$ &142 &-45 &0.44 &0.71675349 \\
139 &GW200311\_115853 &$75.89_{-5.66}^{6.16}$ &3 &-7.5 &1 &-0.93447092 \\
140 &GW200316\_215756 &$25.5_{-1.08}^{8.71}$ &86.5 &47 &1 &0.01299513 \\
141 &GW200322\_091133 &$158.1_{-117.39}^{362.84}$ &60 &17 &0.5 &-0.32705584 \\
142 &GW200322\_091133 &$158.1_{-117.39}^{362.84}$ &200 &17 &0.5 &0.76891174 \\
\\
\hline
\\
\end{longtable}
\endgroup

\bibliographystyle{JHEP}
 \bibliography{GW_refrs}

\end{document}